\documentclass[12pt,preprint]{aastex}

\begin{document}

\title{Gas Giants Formed by Gravitational Instability May Accrete
Atmospheres with Super-Stellar Carbon to Oxygen Ratios}

\author{Alan P. Boss}
\affil{Earth \& Planets Laboratory, Carnegie Institution for Science, 
5241 Broad Branch Road, NW, Washington, DC 20015-1305}
\authoremail{aboss@carnegiescience.edu}

\begin{abstract}

 Characterizing the atmospheric compositions of exoplanets, along with determining 
properties such as their mass, mean density, and orbital configuration, is thought 
to be an effective means for differentiating between various formation and
evolution scenarios. Exoplanet atmospheric C/O ratios, when compared to host 
star C/O ratios, have been advanced as discriminators of gas giant formation 
and evolution scenarios in the context of the core accretion mechanism.
Gas giants formed by gas disk gravitational instability (GDGI), on the other hand, 
are thought to have atmospheres with C/O ratios identical to their host stars.
We examine this assumption through analysis of fully three dimensional radiative
hydrodynamics models of the GDGI in the flux-limited diffusion approximation.
We show here that GDGI protoplanets may be able to form and accrete disk gas with 
super-stellar C/O ratios, as a result of their formation and orbital evolution
in a disk with midplane temperatures in the range of the evaporation temperatures
of water ($\sim$ 135 K) and CO$_2$ ($\sim$ 47 K) ices. Solids that 
avoid fragmentation and grow rapidly to cm-size could be transported
inward to the central protostar or outward to the edge of the
disk considerably faster than the disk gas is dissipated, leading to the
preferential accretion of C-rich disk gas compared to the O-rich ices, provided
that the protoplanet's orbit remains outside $\sim$ 7 au from a solar-mass
protostar. Orbits inside $\sim$ 7 au, however, could result in the accretion
of disk gas with nearly stellar C/O.

\vspace{0.25in}

{\it Unified Astronomy Thesaurus concepts:} Exoplanet atmospheres (487);
Exoplanet formation (492); Extrasolar gaseous giant planets (509); 
Gravitational instability (668); Protoplanetary disks (1300);
Exoplanet atmospheric composition (2021);

\end{abstract}

\section{Introduction}

 One promising means for furthering our understanding of planet formation and evolution 
processes is comparing observations of exoplanet atmospheric abundances with the 
predictions of various planet formation and evolution models 
(e.g., N. Madhusudhan 2019; A. B. T. Penzlin et al. 2024).
K. I. \"Oberg et al. (2011) proposed using planet formation models to explain
differences in the atmospheric abundances of carbon and oxygen for observable 
exoplanets compared to the C/O ratio of their host stars. Giant planet atmospheres
with C/O ratios greater than their host stars, i.e., super-stellar C/O ratios,
could be explained by giant planet formation by the classical core accretion
mechanism (e.g., H. Mizuno 1980; J. B. Pollack et al. 1996) 
when the giant planets accrete most of their
atmospheres from the disk gas at orbital distances beyond the water snowline,
leaving behind O-rich water ice particles. On the other hand, exoplanet
atmospheres that are contaminated by the accretion and evaporation of O-rich 
solids might end up with stellar or even sub-stellar C/O ratios 
(K. I. \"Oberg et al. 2011). Sub-Neptunes, without massive gaseous outer layers,
may have their atmospheric C/O ratios influenced by chemical reactions
with underlying magma oceans, resulting in ratios that range from sub-solar
to super-solar values (Werlen et al. 2025).

 Conventional wisdom holds that gas giants formed by gravitational instability
of the gas in a protoplanetary disk (e.g., A. P. Boss 1997, 2025; R. H. Durisen et al. 2007) 
should lead to atmospheres with C/O ratios similar to that of their host stars. 
Contrary to the expectation that disk instability planets should have nebular composition, 
A. C. Boley et al. (2011) pointed out that heavy-element enrichment in gas giant
exoplanets need not be considered as evidence for their formation by core
accretion. Rather, they showed that gas giants formed by gravitational instability can have 
either sub-nebular or super-nebula heavy-element compositions, the former as a result of 
tidal stripping, and the latter as a result of planetesimal capture and other effects. 
J. D. Ilee et al. (2017) presented a detailed chemical evolution calculation of
fragments formed by GDGI, finding that solids might settle to the fragment cores without
releasing their volatile species, resulting in an non-stellar atmospheric C/O ratio.
The purpose of this paper is to further examine the conventional 
wisdom regarding atmospheric C/O ratios for disk instability protoplanets.

\section{Atmospheric C/O Observations and Models}

 The first transiting exoplanet, the hot Jupiter HD 209458 b, has an atmosphere
with a low C/O ratio of $\sim$ 0.054, in spite of a near-solar metallicity for
its host star (N. Bachmann et al. 2025). The field of exoplanet astronomy thus
began with an object with a puzzling atmospheric C/O ratio. The previous work
cited in this section is an attempt to summarize some of the key findings to date
for both observations and theoretical modeling, in order to further motivate the
present modeling effort.

 J.-B. Ruffio et al. (2021) found similar C/O ratios for the three gas giants
HR 8799 b, c, and d, and argued that these stellar ratios could be a result of
formation by gravitational instability or by pebble accretion along with planetesimal 
enrichment. J. Wang et al. (2023) found a stellar C/O ratio of 0.67 for HR 8799c, 
in spite of the gas giant's super-stellar C/H and O/H abundances, which suggested
rapid formation by gravitational instability within $\sim$ 1 Myr, followed by 
further accretion of CO ices in the outer disk.
P. Molli\`ere et al. (2022) presented detailed models of the expected atmospheric
ratios of C/H and O/H for HR 8799e, finding that models 
of pebble accretion could account for the observed roughly solar C/O ratio of 0.52
for HR 8799e. J. Williams et al. (2025) have considered the effect on pebble 
accretion of ices of mixed composition, e.g., CO ice entrapped in water ice, and shown 
that this effect can lead to disk gas with enhanced C/O inside the water snow line.
On the other hand, B. O’Donovan and B. Bitsch (2026) used pebble accretion to 
predict atmospheric C/O ratios in the range of $\sim$ 0.05 to 0.6 for five 
transiting gas giant exoplanets.

 J. W. Xuan et al. (2024) found by direct imaging that eight substellar companions 
with masses $\sim$ 10-30 $M_{Jupiter}$ all had C/O ratios close to the solar ratio, 
while their host stars had solar abundances. They concluded that formation by direct
gravitational collapse and fragmentation (e.g., A. P. Boss 1981) was likely for
these companions, though core accretion beyond the CO snowline might be possible
in a massive disk with rapid core assembly.
A. Denis et al. (2025) used direct imaging to determine C/O $\approx 0.55$ for
AF Lef b, a gas giant with a mass of $\approx 3.9 M_{Jupiter}$, orbiting at
$\approx 9$ au from its host F8 star, which has a super-solar metallicity.
They cautioned though about drawing any firm conclusion about where the planet
might have formed, if it formed by core accretion.

 T. Molyarova et al. (2025) calculated highly detailed models of the chemical 
composition of disk gas and solid ices with a thin-disk hydrodynamics code.
They considered giant planet formation by both core accretion and gas disk 
gravitational instability (e.g., A. P. Boss 1997), finding that the streaming
instability could lead to disk gas with a wide range (0 to 1.4) of sub-stellar to 
super-stellar C/O ratios, while gravitational instability tended to occur
in regions with stellar (0.34) to super-stellar (0.6) C/O ratios. 
They argued that gravitational instability could likely explain the small range of C/O
ratios observed in directly imaged gas giants compared to transiting planets,
supporting their formation by GDGI.

 N. Madhusudhan et al. (2014) considered both core accretion and
gravitational instability as formation mechanisms for hot Jupiters with super-solar O 
and C abundances, finding that both mechanisms could lead to hot Jupiters with C/O ratios
ranging from solar (C/O = 0.54) to sub-solar (C/O $\sim$ 0.1). 
The chemical homogeneity of binary stars, even of wide binaries 
(e.g., K. Hawkins et al. 2020), is taken as evidence for their formation from the 
collapse and fragmentation of the same molecular cloud core (e.g., A. P. Boss 1981).

\section{GDGI Model Calculations} 

 A new gas disk gravitational instability (GDGI) calculation was performed for this paper. 
The new calculation starts from the final time step of a previously published GDGI model, 
model fldA from A. P. Boss (2021). The advantage of starting with the end point of 
model fldA is that the original model required over 4.5 years of computing time on a
dedicated core of the Carnegie memex cluster, in order to lead to robust fragmentation in a 
calculation with extremely high spatial resolution. The new model (fldAC) relaxes the arbitrary
requirement on the A. P. Boss (2021) models that the disks could not cool to temperatures less 
than that of the disk's initial radial temperature profile, i.e., below 40 K in the case
of model fldA, in order to err on the side of not allowing the disk to become overly
gravitationally unstable. The new model replaces this disk temperature constraint with
a minimum disk temperature of 10 K and an outer disk boundary temperature of 10 K, in
order to simulate the cold environment of a giant molecular cloud complex. With the
new temperature constraints, the new model fldAC was calculated for several
months on the Caltech Resnick cluster, sufficient time to allow the disk temperature to relax
below 40 K where warranted by the local thermodynamics and stabilize, so that disk 
temperatures of interest ($\sim$ 20 K) for CO ice stability (K. I. \"Oberg et al. 2011) 
could be studied.

 Rather than undertake explicit calculations
of the C/O distribution in the disk gas and solids in this exploratory study, this paper 
relies on the approach of the pioneering work of K. I. \"Oberg et al. (2011), who based
their derived C/O ratios for disk gas and solids on approximate temperatures for CO (20 K), 
CO$_2$ (47 K), and H$_2$O (135 K) ice particle evaporation in an assumed static disk. 
We follow this simplified approach here, only for a dynamically evolving disk, by assuming 
that icy solids can evaporate and recondense on sufficiently short timescales.This assumption
can be verified by estimating the rate at which a nominal 1-cm radius water ice particle will
sublimate at a disk temperature of 200 K, well above the evaporation temperature of 135 K
assumed for water ice by K. I. \"Oberg et al. (2011). Following the Hertz-Knudsen equation
approach of W. Guo et al. (2025) for calculating sublimation rates for cometary water
ice from an ice layer 1-cm thick, it is straightforward to estimate that a 1-cm radius ice
particle at a temperature of 200 K should sublimate completely away in a time period of 
about $10^{-3}$ yr, i.e., essentially instantaneously compared to the time scales for the 
dynamical evolution of the present GDGI models, which are of the order of orbital periods.

 As model fldA is more or less a typical GDGI model of the type calculated by the author, it 
serves as an example of the general implications for the C/O ratios expected for disk gas 
accreted by gas giant protoplanets formed by a GDGI around a solar-mass protostar 
(A. P. Boss 1997, 2025). We then examine the results for range of protostar masses,
from 0.1 to 2.0 solar masses, from A. P. Boss (2006, 2011).

 The A. P. Boss (2021) calculations consisted of fully three dimensional gravitational
hydrodynamical models of protoplanetary disks orbiting a solar mass protostar, 
including a complete thermodynamical treatment of the disk gas through detailed
equations of state, as well as radiative transfer in the flux-limited diffusion 
approximation, based on the Rosseland mean opacity of dust grains, water vapor, 
and other sources (see A. P. Boss 1984 for details of the opacity routine). 
The code assumes local thermodynamical equilibrium, and hence a single temperature
for the disk gas, dust grains, and any solid particles within. This assumption is
justified by the high optical depths in the disk midplane in the models. Details 
about the flux-limited diffusion approximation method may be found in A. P. Boss 
(2008) and about other aspects of the EDTONS hydrodynamics code in A. P. Boss 
\& E. A. Myhill (1992).

 A. P. Boss (2021) presented the results of eight GDGI models with varied initial
gas disk midplane temperature profiles, calculated with extremely high spatial
grid resolution, equivalent to $\approx 3.7 \times 10^7$ resolution elements.
Even with this degree of spatial resolution, in some models clumps of disk gas
formed that were so dense as to violate either of two criteria for avoiding spurious
fragmentation, namely the Jeans and Toomre criteria (e.g., A. P. Boss et al. 2000;
A. F. Nelson 2006). When that happens, a calculation is halted and restarted from 
a previously stored time step when neither of these two criteria have been violated. 
The critical high density cell is then drained of 90\% of its mass and momentum, 
which is then inserted into a virtual protoplanet (VP; A. P. Boss 2005), conserving
the total mass and angular momentum of the disk and its embedded VPs. The VPs
are then able to orbit the central protostar, subject to the gravitational field
of the entire system of protostar, disk, and VPs. The VPs are able to accrete disk
gas according to the Bondi-Hoyle-Littleton (BHL) formula (e.g., M. Ruffert \& D. Arnett 
1994) and thus to increase in mass at the expense of the local disk gas.

 Model fldA had formed three VPs by the end of the original calculation, 
while a fourth VP was formed during the extended calculation of model fldAC.
Once created, VPs can be lost from the calculation after they strike either
the inner or outer computational boundaries, at 4 au and 20 au, respectively.
Unlike the self-gravitating clumps that formed in model fldA (A. P. Boss, 2021),
the VPs can only grow and cannot lose mass. In the GDGI models of
R. Matsukoba et al. (2023), a massive clump formed with a mass of 
$\sim 10 M_{Jupiter}$, but lost most of its mass as the fixed computational grid
was unable to follow the contraction of the clump to protoplanetary densities,
though the authors argue that a massive clump should survive in a higher
spatial resolution calculation. The present calculations will show that the VPs gain 
considerable mass by BHL accretion of disk gas.

\subsection{Results for new Model fldAC} 

 Model fldAC began with the final time step of model fldA, after 188 yrs of evolution.
The original model fldA
(A. P. Boss 2021) began with an initial midplane disk temperature profile 
that dropped smoothly from 600 K at the inner active computational grid boundary of 
4 au to a uniform outer disk temperature of 40 K beyond about 6 au, and extending to the 
outer disk boundary at 20 au. The initial disk had a mass of 0.091 $M_\odot$. 
Given the initial temperature distribution, the initial disk for model fldA
had a Toomre $Q$ profile that drops to 1.3 in the outer disk, indicating that
the disk is marginally unstable to the growth of nonaxisymmetric features,
such as spiral arms and dense clumps; i.e., the disk is subject to the GDGI.
As is to be expected, the small (1\%) initial nonaxisymmetric density perturbation
to the gas disk grew rapidly and led to the formation of numerous dense spiral
arms and clumps, along with three VPs after 188 yrs of evolution. 

 At this time, the EDTONS code was changed to allow the disk temperature to
drop as low as 10 K, as noted above, and calculation of model fldAC began,
starting from the last timestep of model fldA (A. P. Boss 2021).
A fourth VP formed in model fldAC after just 3.4 yrs of further evolution.
Figure 1 depicts the midplane gas density distribution for model fldAC at the final
time calculated, i.e., after another 8 yrs of evolution ($\sim$ an orbital period at
the inner boundary), showing the dense spiral arms and the four VPs in the disk.
Figure 2 presents the midplane temperature distribution for model fldAC at
the same time as the density distribution in Figure 1. It can be seen that
the disk is threaded by a large number of spiral temperature maxima associated
with the density spiral arms.
Figure 3 plots the midplane temperature profile along an azimuth at 9 o'clock 
in Figures 1 and 2, showing the detailed structure of the temperature distribution 
made possible by the extremely high spatial resolution of the calculation.
Figure 3 is labelled with the key temperatures for CO (20 K), 
CO$_2$ (47 K), and H$_2$O (135 K) ice particle evaporation from Table 1 and Figure 1 
of K. I. \"Oberg et al. (2011). The disk gas is thus effectively finely threaded with 
spiral arm snowlines separating regions with disk gas C/O ratios of 
$\sim$ 1.0 or $\sim$ 0.85, signifying the snowlines for CO$_2$ and H$_2$O, respectively.
Some regions have even dropped below 20 K and hence have the CO locked up in solid
grains. These extremely cold regions with temperatures as low as 10 K are
associated with disk regions with extremely low vertical optical depths, as
shown in the corresponding azimuthal profile in Figure 4, where the midplane is allowed 
to cool down to the assumed background GMC temperature of 10 K.

 The masses and orbital parameters (semimajor axes and
eccentricities) of the four VPs at the final time of  196 yr are listed in Table 1. 
Their final masses ranged from a newly formed VP with a mass of 0.26 $M_{Jupiter}$ to a VP with a 
mass of 1.3 $M_{Jupiter}$. These protoplanets thus have masses that fall in the range between 
Saturn-mass and Jupiter-mass. The VPs can be expected to continue to gain mass by 
further gas disk accretion. Figure 5 explicitly shows the temperatures of the
disk gas accreted by VP2 during the evolution of model fldAC, implying the accretion of 
nearly stellar to super-stellar C/O ratio disk gas with C/O ratios varying 
from $\sim$ 0.6 to $\sim$ 0.85. Two of the other three VPs (VP1 and VP4) similarly 
accrete disk gas with a range of gas temperatures, both above and below the water ice
snowline temperature of $\sim$ 135 K, while VP3 accretes only gas hotter than
135 K during the evolution of model fldAC, due to its orbital semimajor axis being
the smallest of the four VPs (see Table 1). 
As convective mixing in gaseous protoplanets requires high internal temperatures 
($\sim 2 \times 10^{4}$ K; A. Vazan et al. 2018), the atmospheres of gas giant
protoplanets formed by a GDGI beyond about 7 au are likely to have super-stellar C/O, 
regardless of subsequent protoplanet contraction and evolution.

 We now turn to the question of the decoupling of the disk solids from the disk gas
in model fldAC, which is necessary for the present scheme of accumulating 
super-stellar C/O disk gas without the accretion of sufficient sub-stellar C/O ice
particles to restore the bulk gaseous protoplanet to the stellar C/O ratio.
Following A. P. Boss et al. (2020), we define the particle stopping time for cm-sized
solids $t_s$ (Weidenschilling 1977) as 
\begin{equation}
t_s = {\rho_p r_p \over \rho c_s},
\end{equation}
\noindent
with a nominal particle density $\rho_p =$ 3 g cm$^{-3}$, particle radius $r_p =$ 1 cm, 
gas density $\rho$, and sound speed $c_s$. For 1 cm particles, the Epstein drag 
law is usually appropriate for these disk models (Boss et al. 2012).
The Stokes number $St$ is then defined as the 
product of the stopping time $t_s$ and the local disk angular velocity $\Omega$.
A Stokes number $St = 2 \pi$ means the particle stopping time is equal to the orbital period.
Large values of $St >> 1$ imply a particle stopping time of many orbital periods and
hence a particle large enough to not be responsive to gas 
drag, while small values of $St << 1$ imply a particle that is locked to the gas motions.

 Figures 6 (at 188 yrs) and 7 (at 197 yrs) show that the Stokes number varies
drastically throughout the disk midplane, while not varying as strongly as a function
of time. The minimum and maximum values of $St$ found at a given orbital radius can
vary by factors of up to five orders of magnitude, depending on the azimuthal angle
in the disk midplane. Close to the inner edge of the active computational volume
at a radius of 4 au, $St$ can vary from as high as $\sim 10^2$ to as low as
$\sim 10^{-3}$. At the outer edge at 20 au, in contrast, the profiles converge
to $St \sim 0.3$. Relatively large values of 
$St$ occur when particles are embedded in relatively low gas density regions (see Figure 1)
while small values correspond to relatively high gas density. 
Figures 6 and 7 demonstrate that throughout the evolution of model fldAC cm-sized
particles should be strongly decoupled from the gas in portions of the innermost disk, and
hence lost to the growing central protostar, while to a lesser extent, particles
in the outermost disk will be decoupled and tossed further outward
by the gravitational tugs of the spiral arms. Note that
during the course of the model fldA and model fldAC evolutions, the central protostar
accreted $\sim 0.01 M_\odot \sim 10 M_{Jupiter}$ of material from the inner disk.
 
 This Stokes number analysis for cm-sized solids is similar to that seen in 
GDGI previous models (A. P. Boss et al. 2020) that led to the conclusion that cm-sized
particles should be depleted in the gas giant planet-forming regions of the disk.
Further details regarding this key point are presented in the next section, but the
point is that dense spiral arms can transport cm-sized particles both inward and outward
to regions where they can decouple from the gas disk and be lost by evaporation 
in the inner disk or tossed outward beyond the gas giant planet formation region.

 Stokes numbers in the wide range of values from $\sim 10^{-4}$ to $\sim 10$
have been inferred for seven protoplanetary disks using ALMA/VLA observations 
(H. Jiang et al. 2024), over scales of $\sim 10$ au to $\sim 200$ au (see their
Figure 3). These values are roughly consistent with those presented in Figures 6 and 7,
though the present models extend only out to 20 au. H. Jiang et al. (2024) also
found that the maximum pebble sizes inferred from their theoretical modeling
ranged from $\sim 0.1$ cm to $\sim 1$ cm for six disks, to as large as $\sim 10$ cm
for TW Hya (their Figure 2). This analysis appears to support the basic assumption 
here that dust and ice grains can grow to cm-size in protoplanetary disks.

 The fact that cm-sized refractory inclusions have been found in Comet Wild 2
(D. J. Joswiak et al. 2017) requires the large scale transport of cm-sized solids
from the inner solar nebula to the outer regions where comets form, transport which
has been shown to occur during GDGI in the FU Orionis phase of protostellar evolution 
(A. P. Boss et al. 2020). This gives strong observational support for the assertion here that 
cm-sized ice particles with sub-stellar C/O should be depleted in the gas giant planet
formation region, leading to preferential accretion of super-stellar C/O disk gas
by newly formed gas giant protoplanets. A protoplanet with a bulk composition with
a super-stellar C/O ratio is likely to have a super-stellar C/O atmosphere, regardless
of the subsequent contraction and evolution towards planetary densities.

\subsection{Results for Other GDGI Models} 

 The present calculation focuses on the disk temperature distribution in a GDGI 
unstable disk that forms gas giant protoplanets around a solar-mass protostar.
The question then arises if the results are limited to solar-mass stars.
We thus now turn to the results for gas giants formed by GDGI around protostars of
a wide range of masses (A. P. Boss 2006, 2011). These GDGI models employed the
same three dimensional radiative hydrodynamics code as was used in model fldA
(A. P. Boss 2021). A. P. Boss (2006) presented the results of eleven GDGI models 
with initial protostar masses of either 0.1 or 0.5 $M_\odot$, i.e., M dwarfs.
The initial disks had masses ranging from 0.021 to 0.065 $M_\odot$, minimum
Toomre $Q$ values of 1.40 to 1.63, and outer disk temperatures from 20 K to 60 K.
Most of the 0.5 $M_\odot$ protostar models fragmented into long-lived dense 
clumps with Jupiter-like masses, surrounded by disk gas with a temperature of 
$\sim$ 70 K (Figure 5 of A.P. Boss 2006). Several of the 0.1 $M_\odot$ protostar
models also formed Jupiter-mass dense clumps, though embedded in lower temperature
disk gas at $\sim$ 40 K. 

 A. P. Boss (2011) showed results for five GDGI models with protostar masses of
0.1, 0.5, 1.0, 1.5, and 2.0 $M_\odot$. These models were intended to explore
the formation of gas giants by GDGI at larger distances than the A. P. Boss (2006, 2021)
models, and so started with disks with outer radii of 60 au and masses ranging
from 0.028 to 0.21 $M_\odot$, respectively. All started from highly unstable
minimum Toomre $Q$ values of $\sim 1.1$, and hence all rapidly fragmented into
two to three dense clumps with masses from 0.8 to 4.9 $M_{Jupiter}$. The spiral
arms in the disk gas distributions had peak temperatures ranging from
$\sim$ 40 K to $\sim$ 70 K, respectively. Hence these models again demonstrate
that gas giants formed by a GDGI should accrete super-stellar C/O disk gas.

\section{Discussion}

 The implications for exoplanet super-stellar C/O atmosphere ratios derived here require that
gas giants formed by GDGI in a disk where a significant fraction (by mass) of the dust
and ice grains have grown to $\sim$ cm-size and have been removed from the orbital
locations of the protoplanets, allowing the gas giant protoplanets to accrete disk gas 
with super-stellar C/O. We now examine each of these requirements in order.

\subsection{Recent Observations Supporting Rapid Gas Giant Planet Formation}

 A high-resolution survey of 90 protostellar disks found the presence of disk 
substructures in even the earliest protostellar stage disks (C.-H. Hsieh et al. 2025),
suggesting that the formation of planetesimals and of giant planets occurs earlier 
and more efficiently than is expected in the core accretion paradigm 
(e.g., J. B. Pollack et al. 1996). Their study pointed to planet formation
beginning in the earliest stages of protostellar evolution, while the central
protostar is still accreting mass at a high rate. GDGI satisfies the rapid
formation time scales implied by this recent survey, as a GDGI proceeds on the
time scale of a few gas disk orbital periods (e.g., A. P. Boss 1997).

\subsection{Observed Grain Sizes in Protoplanetary Disks}
 
 Radio wave observations have shown that dust grains range in size from the 
sub-micron in the interstellar medium to as large as cm-size in protoplanetary 
disks (e.g., L. Cacciapuoti et al. 2025). The PMS star CQ Tau has been shown to 
have dust grains as large as a few cm in its circumstellar disk (L. Testi et al. 2003).
Radio wave observations of the compact dust ring surrounding the Class I protostar WL17 
imply that the dust grains have grown as large as 4.2 mm, while other protostars
have been suggested to have cm-sized grains (J. Hashimoto et al. 2025).

\subsection{Grain Growth to cm-Size}

 E. I. Vorobyov et al. (2025) studied the earliest phases of protoplanetary disk formation
with a hydrodynamics code that followed both the collapsing molecular cloud core gas
and the growth of the embedded dust grains by collisions and coagulation. They found
within 300 yr after the formation of the disk, dust grains had grown significantly
throughout the extent of the disk, of radius $\sim$ 10 au (their Figure 7).
Furthermore, the models show that the maximum dust grain size after 12 kyr is
$\sim$ 100 cm at $\sim$ 1 au and $\sim$ 1 cm at $\sim$ 10 au when grains are not
prevented by fragmentation to growing so large (their Figure 3). These results
suggest that the presence of significant numbers of cm-sized solid particles is
to be expected during the earliest phases of protoplanetary disk evolution.

 However, it should be noted that the calculations of E. I. Vorobyov et al. (2025)
did not allow for fragmentation of particles following collisions. 
Theoretical studies of turbulent gas disks have been used to approximate the
relative velocities expected for grains of varying sizes (e.g., C. W. Ormel \&
J. N. Cuzzi 2007; T. Birnstiel et al. 2011), the latter finding that ``relative velocities
increase with grain size'', potentially leading to further growth stalling at
some particular grain size, depending on the disk properties, such as density,
temperature, and strength of turbulence. The predictions of the present paper are 
based on the assumption that grains can grow to at least cm-size.

\subsection{Fate of cm-Size Grains in a MGU Disk}

 A. P. Boss (2015) studied the orbital evolution of cm-sized to 10-m-sized particles
embedded in a marginally gravitationally unstable (MGU) protoplanetary disk.
While most of the 1 to 10 m particles survived in the outer disk over time scales
of $\sim 10^4$ yr, most of the 1 and 10 cm particles suffered inward migration and
accretion onto the central protostar, as smaller particles are more closely tied
to the disk gas. Typically only 5\% or less of the cm-sized particles survived these
interactions with the MGU disk gas by escaping to the outer edge of the disk.
A. P. Boss et al. (2020) modeled the three-dimensional orbital evolution of cm-sized 
calcium-aluminum-rich inclusions (CAIs) during the FU Orionis outbursts 
(e.g., L. Hartmann \& S. J. Kenyon 1996) thought to characterize
the earliest evolution of young solar-type stars. Their models allowed the particles
to traverse trajectories that were not limited to the disk midplane, as was the case
with the A. P. Boss (2015) models. The particles could settle rapidly to the 
disk midplane, only to be lofted upward in the disk by periodic encounters with
the shock fronts in the MGU disk gas. Rapid transport of the particles both inward
onto the central protostar and outward to the edge of the disk occurred within
$\sim 10^4$ yr, resulting in the loss of over 95\% of the cm-size particles
at the expense of the loss of $\sim$ 60\% of the disk gas onto the protostar.
This implies that the ratio of remaining cm-sized particles to remaining disk 
gas mass has dropped by a factor of about 10 from that of the initial disk 
as a result of gas-solid decoupling in the A. P. Boss et al. (2020) MGU disk models.
Combined with the lofting of the remaining particles up and away from protoplanets 
orbiting in the disk midplane, cm-size particle accretion by a growing protoplanet
is likely to be extremely limited. These models imply that the remaining gas in a 
MGU protoplanetary disk should have super-stellar C/O, given the decoupling 
with cm-sized ice and dust grains.

 The models of E. I. Vorobyov et al. (2025) support the basic findings of A. P. Boss 
(2015) and A. P. Boss et al. (2020) regarding the decoupling of cm-sized solids from the 
disk gas during MGU disk phases. Their Figure 10 shows that when cm-sized solids are 
allowed to form, by $\sim 10^4$ yr the total dust to gas surface density is enhanced in 
portions of both the innermost and outermost regions of the disk by factors of up to 10 
compared to the initial abundance ratio of 1 to 100, while when m-sized solids formed, 
this resulted in much of the dust being transported to the outermost disk and beyond at the 
expense of the innermost disk, consistent with the above findings of A. P. Boss (2015).
E. I. Vorobyov et al. (2025) state that dust enhancements are ``highly nonhomogeneous 
across the disk. Local enhancements in dust alternate with depletions.'' This is consistent with 
the huge variations in Stokes numbers with azimuthal angle that occur in the
present models (Figures 6 and 7) and by the spiral clumps and voids evident in Figure 1
and in Figure 8 of A. P. Boss (2015).
The dust grain densities in the E. I. Vorobyov et al. (2025) models (their Figure 2) 
show variations by up to factors of about 10 in azimuth at any given radial distance, 
consistent with the ``nonhomogeneous'' azimuthal midplane density plots in their Figure 10.

 E. I. Vorobyov et al. (2025) conclude that local enhancements are ``likely due to 
local processes within the disk, such as the radially varying efficiency of vertical 
dust settling and perturbations by spiral density waves.'' Figure 10 
of E. I. Vorobyov et al. (2025) shows a good correlation of the spiral arms with
the regions where the dust to gas surface density is close to the initial ratio; for the
cm-size model, where spiral arms exist, they delineate regions at the same orbital distance
with either enhanced or depleted dust to gas surface densities. Outside of the radius of the
region where spiral arms are specified, the dust to gas ratio is nearly uniformly enhanced 
over the initial ratios, as it is inside the orbital radius with the specified spiral arms. 
Clearly the spiral arms are heavily
involved in producing the ``nonhomogeneous'' dust to gas ratio regions seen in
the models of E. I. Vorobyov et al. (2025). Figures 11 and 16 of A. P. Boss et al. (2020)
showed that significant vertical mixing also occurs in these prior models, along
with spiral arms, which first suggested that cm-sized solids would be removed from 
the disk faster than the disk gas (see Figure 9 of A. P. Boss et al. 2020).

 J. D. Ilee et al. (2017) reached a similar conclusion as the present work regarding 
the possibility of non-stellar atmospheric C/O ratios for gas giants formed by GDGI, but
based on a different physical mechanism, namely the rapid settling of icy solids to the
centers of the protoplanets, without loss of their volatile components to the outer regions. 
Here we propose a second mechanism, namely large-scale transport of $\sim$ cm-sized solids 
both inward and outward from the gas giant planet formation region.

\subsection{Gas Giant Orbital Evolution and Disk Gas Accretion}

 The GDGI is capable of forming gas giant protoplanets on initial orbits ranging
from semimajor axes of $\sim$ 1 au to $\sim$ 100 au (e.g., A. P. Boss 2024, 2025).
A. P. Boss (2013) studied the orbital migration of giant planets formed by either core 
accretion or by a GDGI by following the orbital evolutions of virtual protoplanets (VPs)
undergoing gravitational interactions with each other, the central protostar, and
a MGU disk. That study showed that giant
protoplanets initially distributed between 6 au and 12 au in a MGU disk could orbit
relatively stably with semimajor axes $\sim$ 10 au without undergoing monotonic 
inward (or outward) orbital migration. Instead, the protoplanets continued to gain
mass by disk gas accretion, at rates of $\sim 10^{-4} M_{Jupiter}$ yr$^{-1}$,
while suffering moderate orbital eccentricities as a
result of gravitational interactions with the spiral arms in the MGU disk. A MGU-disk 
phase is a likely explanation (e.g., Z. Zhu et al. 2010; E. I. Vorobyov \& S. Basu 2010)
for the FU Orionis phase of protostar evolution.

\section{Conclusions}

 The seminal paper on C/O atmospheric ratios by K. I. \"Oberg et al. (2011) has led 
to a number of applications of their basic concept in the context of the core accretion
mechanism for giant planet formation, applications of increasing complexity and detail. 
The motivation for the present paper is to build upon the basic concept of the
K. I. \"Oberg et al. (2011) paper now in the context of the GDGI mechanism for gas
giant formation. While many details remain to be studied in the GDGI formation context 
(e.g., stalling of grain growth by collisions resulting in fragmentation,
further orbital evolution with accretion of solids and disk gas, contraction of 
the gaseous protoplanet, core formation, internal mixing and differentiation,
atmospheric stability, etc.), the goal of this paper was simply to demonstrate that
gas giants formed by GDGI need not necessarily have the atmospheres with solar or 
stellar C/O ratios that are commonly assumed to be the likely outcome. This results from
the possibly more rapid inward and outward orbital transport of sub-stellar C/O 
cm-sized ice particles 
in a MGU disk compared to the super-stellar C/O disk gas. For a range of protostar masses, gas 
giants that form and continue to orbit beyond $\sim$ 7 au are likely to accrete disk 
gas with C/O $\ge$ $\sim$ 0.85, while those that form and orbit inside this radius are
likely to accrete disk gas with C/O $\sim$ 0.6, i.e., gas with nearly stellar C/O.
The degree to which these ratios are non-stellar will depend on the degree to which 
ice particles of various sizes are lost to the inner or outer disk, which will require
further GDGI gas and particle dynamics calculations in order to make a firmer prediction.

 The original computations were performed on the Carnegie Science memex computer 
with the support of the Carnegie Scientific Computing Committee. 
The new calculations were performed on the Resnick High Performance Computing 
Center at Caltech with the support of the Carnegie Institution for Science.
The paper has been considerably improved as a result of several sets of comments provided
by the reviewer.

\clearpage
\begin{deluxetable}{lccccc}
\tablecaption{Gas giant protoplanet properties at the termination of model fldAC (a continuation of model fldA of A. P. Boss 2021):
VP number, mass (Jupiter masses), semimajor axis (au), orbital eccentricity, local VP disk temperature (K). }
\label{tbl-1}
\tablewidth{0pt}
\tablehead{\colhead{VP \#}
& \colhead{$M_p$}
& \colhead{$a$}
& \colhead{$e$}
& \colhead{$T_{VP}$}}
\startdata

VP1 & 0.48  & 5.9 & 0.28 & 250 \\

VP2 & 0.64  & 6.6 & 0.087 & 133 \\

VP3 & 1.3   & 4.8 & 0.16 & 275 \\

VP4 & 0.27  & 5.7 & 0.32 & 150 \\  

\enddata
\end{deluxetable}

\clearpage

\begin{figure}
\vspace{-2.0in}
\plotone{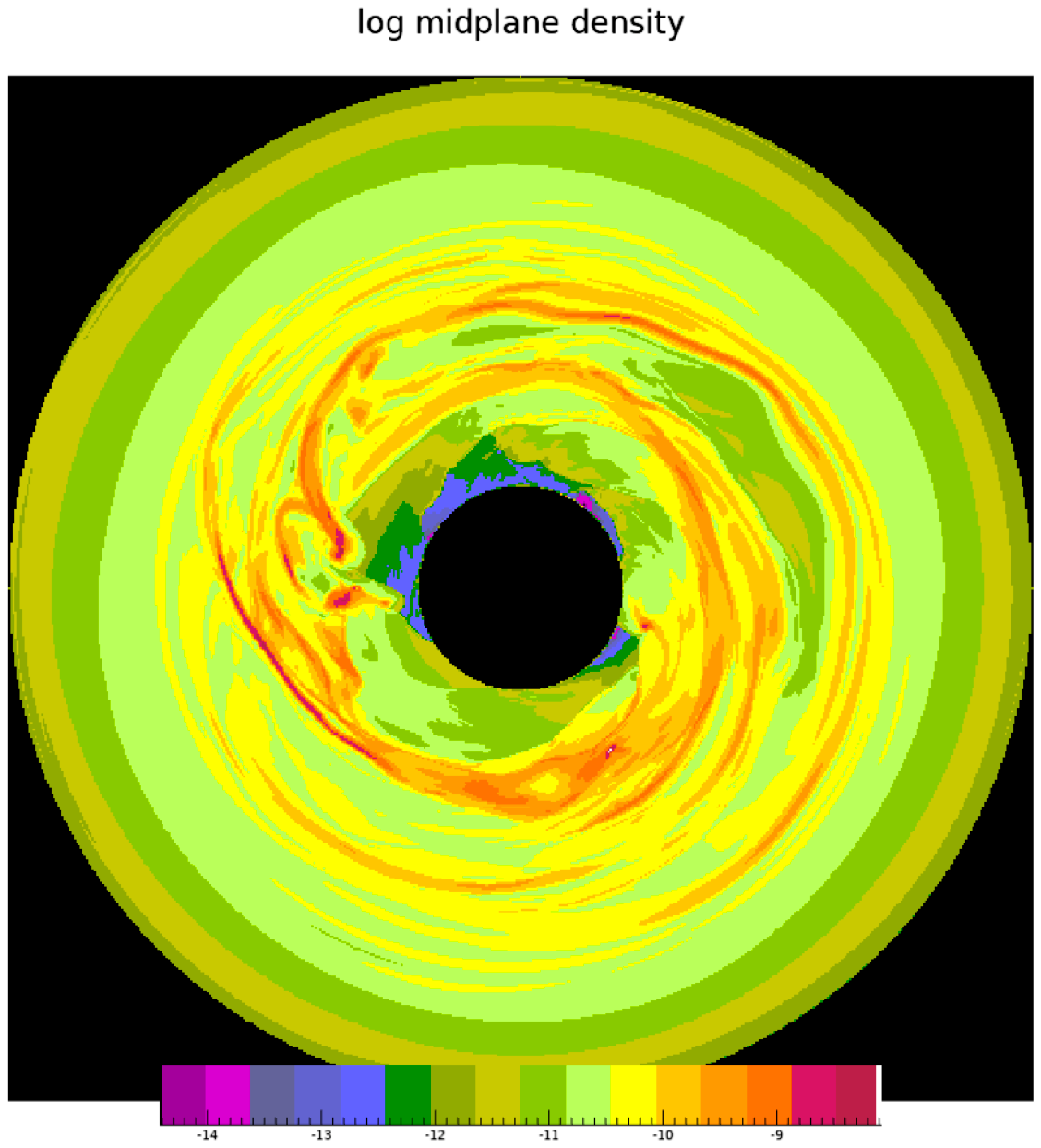}
\vspace{0.0in}
\caption{Log midplane density (g cm$^{-3}$) contours for model fldAC after 196
yr of evolution. Numerous dense spiral arms and the four virtual protoplanets (VPs) 
are evident as dark red features in the inner disk. 
The disk has an inner radius of 4 au and an outer radius of 20 au.}
\end{figure}

\clearpage

\begin{figure}
\vspace{-2.0in}
\plotone{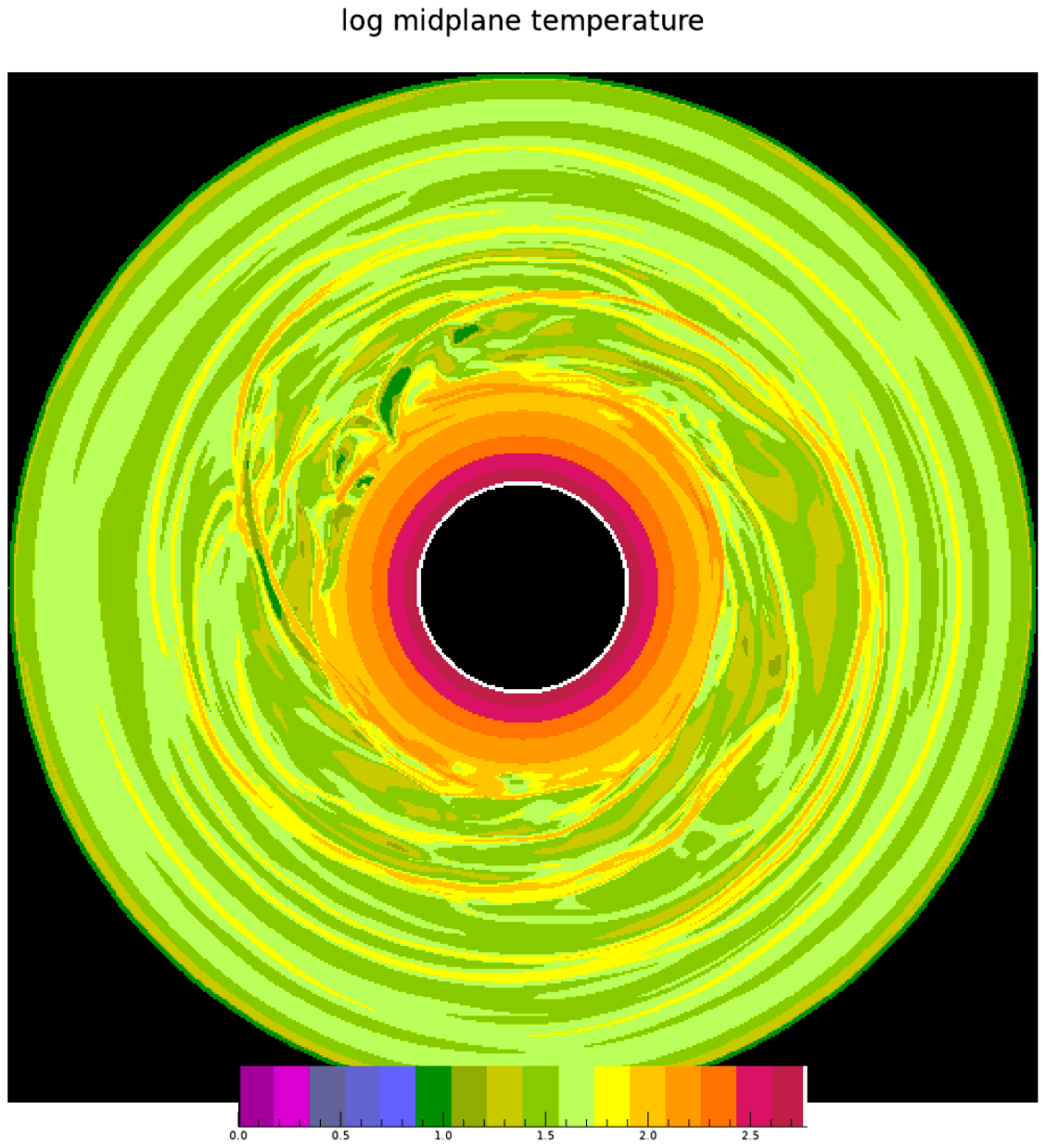}
\vspace{0.0in}
\caption{Log midplane gas temperature (K) contours for model fldAC, same as in Figure 1.}
\end{figure}

\clearpage

\begin{figure}
\vspace{-2.0in}
\plotone{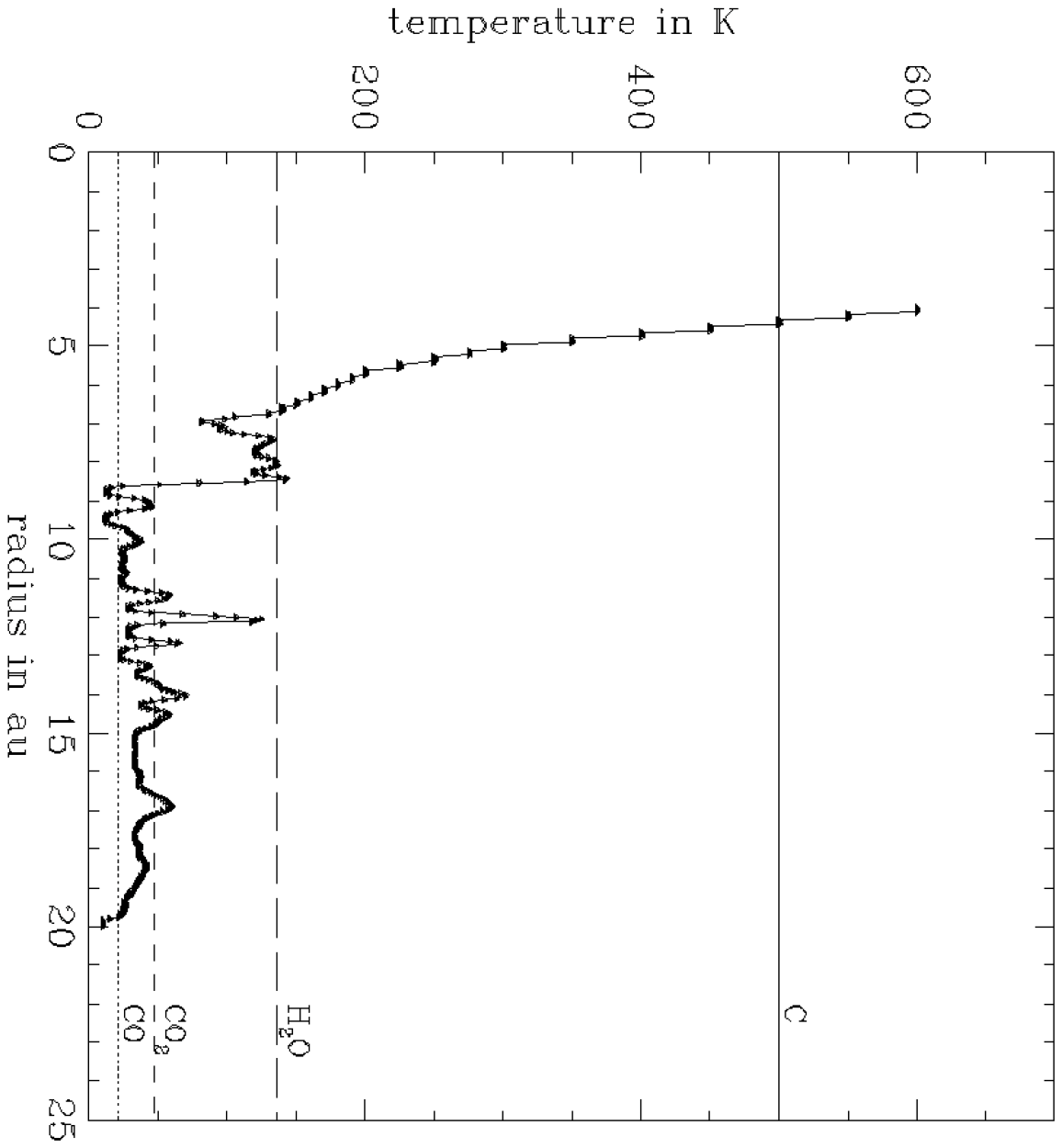}
\vspace{0.0in}
\caption{Midplane temperature profile along an azimuth at 9 o'clock in Figures 1 and 2
for model fldAC. Snowlines for C, H$_2$O, CO$_2$ and CO are shown. 
The profile passes through regions of different gas C/O ratios based on the disk gas
temperatures (as in K. I. \"Oberg et al. 2011): $C/O \approx 1.0$ ($T < 47 K$), 
$C/O \approx 0.85$ ($135 K > T > 47 K$), and $C/O \approx 0.6$ ($T > 135 K$).}
\end{figure}

\clearpage

\begin{figure}
\vspace{-2.0in}
\plotone{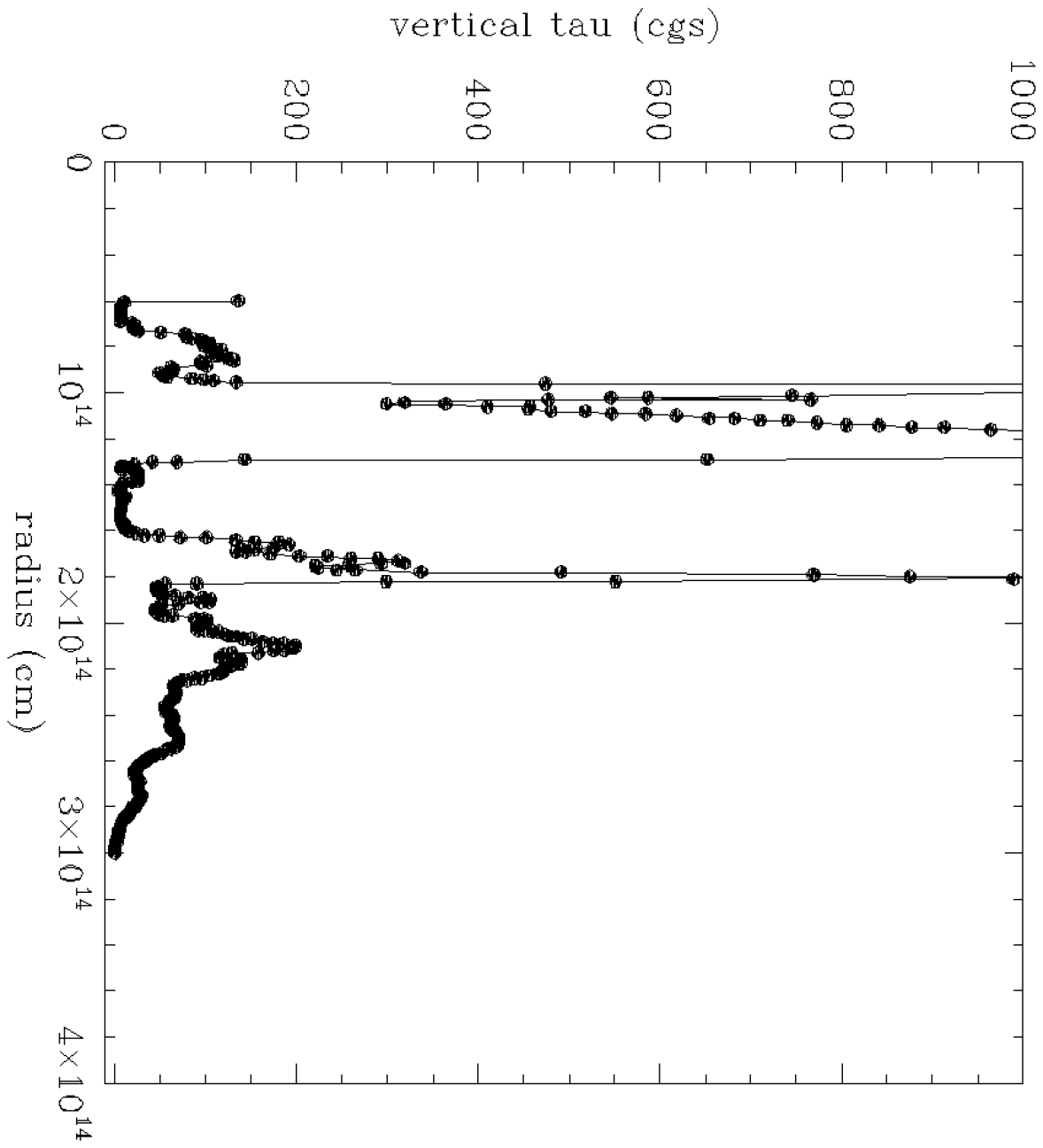}
\vspace{0.0in}
\caption{Vertical optical depth from the disk surface to the disk midplane along 
the same azimuth as in Figure 3 for model fldAC. Low optical depths at some radii
are responsible for rapid disk cooling toward the background GMC temperature
of 10 K.}
\end{figure}

\clearpage

\begin{figure}
\vspace{-2.0in}
\plotone{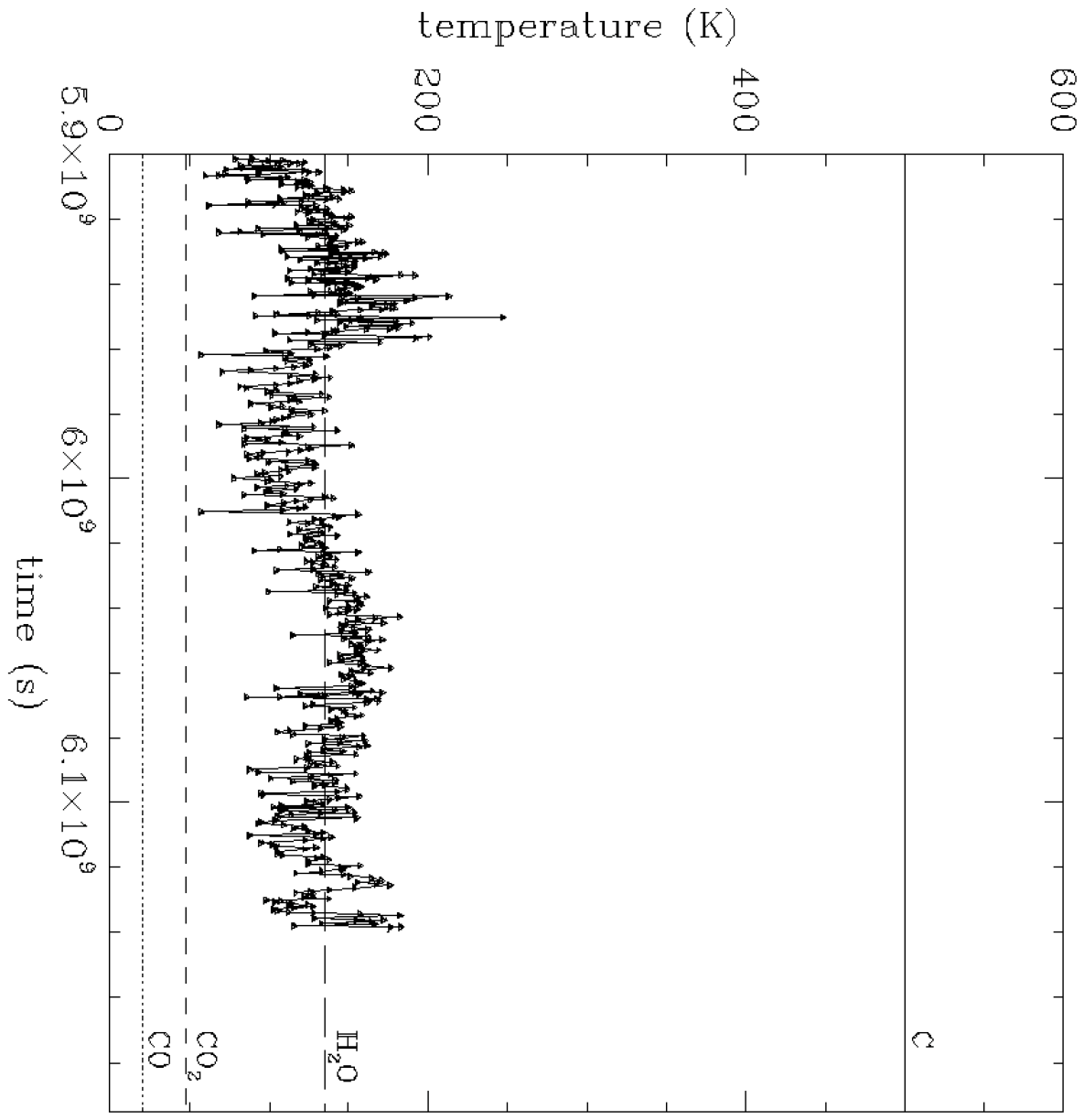}
\vspace{0.0in}
\caption{Temperatures of the local disk gas accreted by VP2 during the evolution of
model fldAC as it gains mass by BHL accretion. Snowlines for C, H$_2$O, CO$_2$ 
and CO are shown, implying the accretion of disk gas with $C/O \approx 0.85$ when $135 K > T > 47 K$
and $C/O \approx 0.6$ when $T > 135 K$.}
\end{figure}

\clearpage

\begin{figure}
\vspace{-2.0in}
\plotone{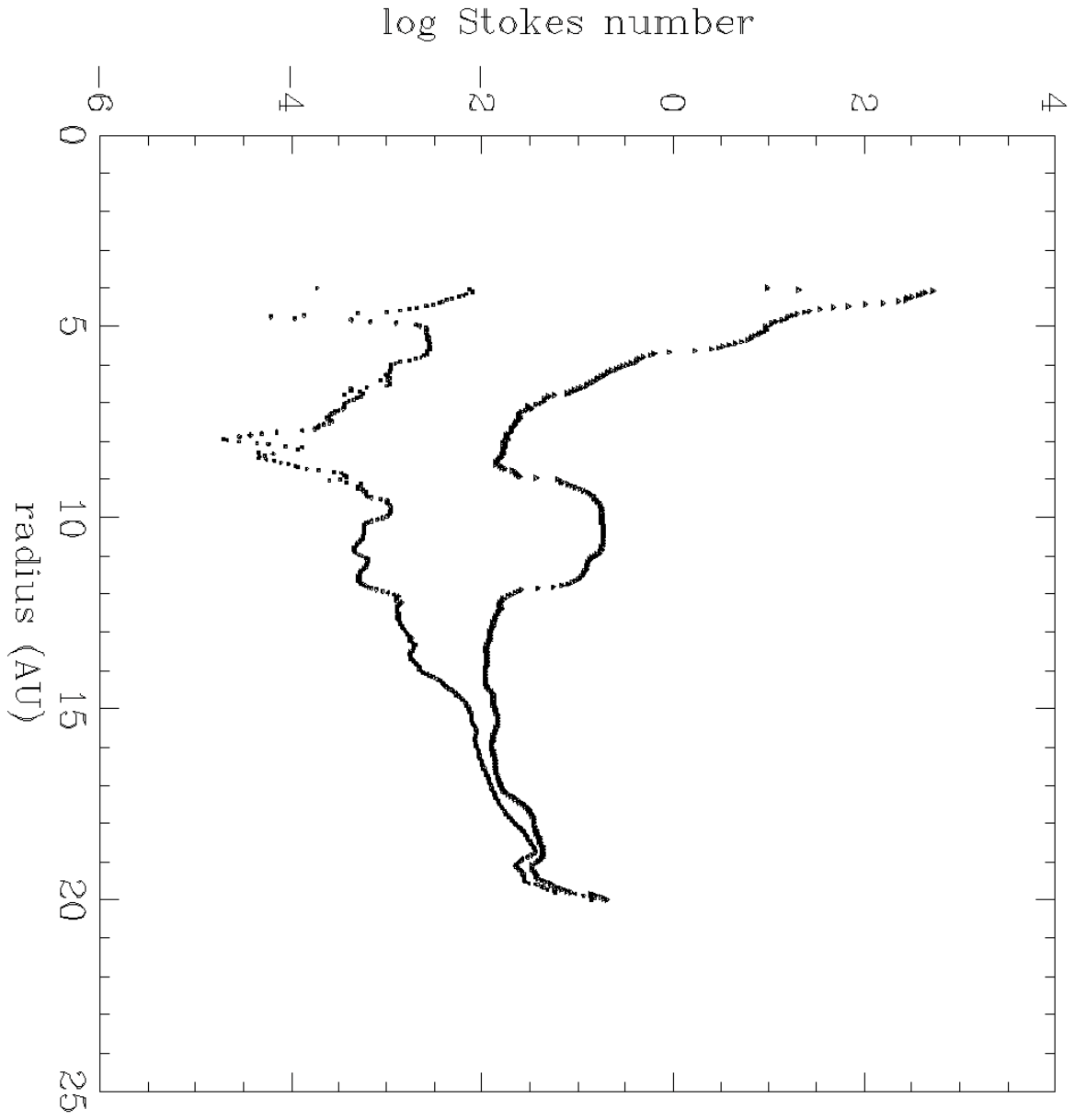}
\vspace{0.0in}
\caption{Radial profile of the Stokes number of 1-cm-size particles orbiting in the 
midplane of model fldAC after 188 yrs of evolution, showing the maximum (top) and
minimum (bottom) value of the Stokes number for the entire azimuthal grid.}
\end{figure}
\clearpage

\begin{figure}
\vspace{-2.0in}
\plotone{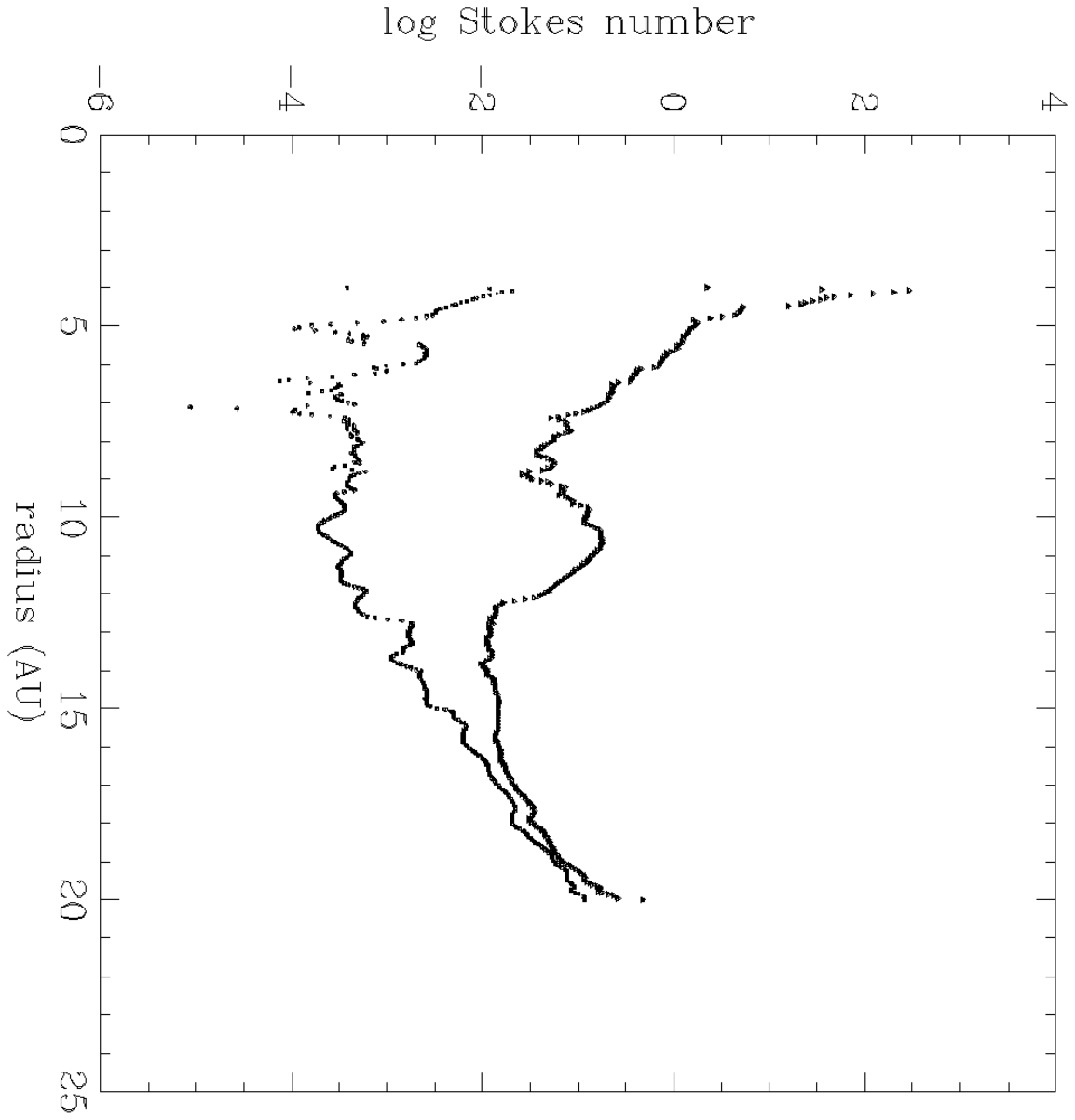}
\vspace{0.0in}
\caption{Same as Figure 6 except at the later time of 196 yrs of evolution.}
\end{figure}

\end{document}